\documentclass[11pt,a4paper]{article}
\pdfoutput=1
\usepackage{graphicx}
\usepackage{jheppub}

  \usepackage{amsmath}
  \usepackage{amsfonts}
  \usepackage{amssymb}

  \usepackage{graphicx}
  \usepackage{float}
  \DeclareGraphicsExtensions{.ps}

\title{Simulations of Cold Electroweak Baryogenesis: Finding the optimal quench time}

	\author[b]{Zong-Gang Mou,}
	\author[a]{Paul M. Saffin,}
	\author[b]{Anders Tranberg}

	\affiliation[a]{School of Physics and Astronomy, University Park, University of Nottingham,\\ Nottingham NG7 2RD, United Kingdom}
	\affiliation[b]{Faculty of Science and Technology, University of Stavanger, 4036 Stavanger, Norway}

	\emailAdd{zonggang.mou@uis.no}
	\emailAdd{paul.saffin@nottingham.ac.uk}
	\emailAdd{anders.tranberg@uis.no}
	

\abstract{We revisit the numerical computation of the baryon asymmetry from Cold Electroweak Baryogenesis given the physical Higgs mass. We investigate the dependence of the asymmetry on the speed at which electroweak symmetry breaking takes place. The maximum asymmetry does not occur for arbitrarily fast quenches, but at quench times of about $\tau_q\simeq 16\, m_H^{-1}$, with no asymmetry created for quenches slower than $\tau_q>30\, m_H^{-1}$. Curiously, we also find that the overall sign of the asymmetry depends on the quench time. }

\begin{document}

\maketitle

\section{Introduction}
\label{sec:Intro}

Cold Electroweak Baryogenesis (EWBG) was proposed some time ago as an alternative mechanism for generating the observed baryon asymmetry of the Universe \cite{Krauss:1999ng,GarciaBellido:1999sv,Copeland:2001qw}. The original (Hot) Electroweak Baryogenesis relies on a strong finite temperature electroweak phase transition to provide the required out-of-equilibrium conditions \cite{Kuzmin:1985mm,Cohen:1993nk}. In the Cold scenario, electroweak symmetry breaking occurs through a spinodal decomposition from an initial state at zero (or very low) temperature.

Cold EWBG hence sidesteps the requirement of a first order phase transition, which is absent in the minimal Standard Model (at the physical Higgs mass), and increasingly constrained by experiment in simple extensions (for recent work, see for instance \cite{Damgaard:2013kva,Haarr:2016qzq,Beniwal:2017eik}). The scenario is very simple, and allows for straightforward first-principles numerical simulations of the entire baryogenesis process. The Hot scenario in contrast is a quite complex sequence of events separated in space and time. 

Challenges of the Cold scenario include the origin of the cold initial state and the subsequent quench, and the origin of the required CP-violation. The former is typically ascribed to a coupling of the Higgs field to another scalar, whose dynamics triggers the symmetry breaking quench \cite{Copeland:2001qw,Enqvist:2010fd,vanTent:2004rc,Konstandin:2011ds}. The resulting baryon asymmetry has been computed for a number of implementations and model choices, notably extensions of the Standard Model with additional scalar fields and CP-violation \cite{Rajantie:2000nj,Tranberg:2003gi,Tranberg:2006ip,Tranberg:2006dg,Tranberg:2012jp,Tranberg:2012qu,Mou:2015aia}. 

From a model-building perspective, it is important to determine, as model-independently as possible, the dependence of the asymmetry on the speed at with the spinodal quench is performed. All things being equal, one might expect that a fast quench gives a state further from equilibrium, and therefore a larger asymmetry. In an exploratory publication \cite{Tranberg:2006dg}, it was found that indeed fast quenches give an asymmetry while slow ones did not, quantified through the speed at which the Higgs mass parameter "flipped" in the potential,
\begin{eqnarray}
V[\phi] = \mu^2(t)\phi^\dagger\phi + \lambda (\phi^\dagger\phi)^2,
\end{eqnarray}
with for $t<\tau_q$
\begin{eqnarray}
\mu^2_{\rm eff}(t) = \mu^2\left(1-\frac{2t}{\tau_q}\right), 
\end{eqnarray}
and  for $t>\tau_q$, $\mu_{\rm eff}^2(t)=-\mu^2$. One may define a characteristic quench speed as
\begin{eqnarray}
u=-\frac{\sqrt{2}}{m_H^3}\frac{d\mu^2_{\rm eff}(t)}{dt}|_{\mu_{\rm eff}=0}=\frac{\sqrt{2}}{m_H \tau_q}.
\end{eqnarray}
Results indicated that quenches slower than $u\simeq 0.1$, $m_H\tau_q>10-15$ are too slow to generate an asymmetry. But because of the vast numerical effort required, and because the results were very sensitive to the (at the time) unknown value of the Higgs mass, no further progress was made. 

In the present paper, we revisit this computation, using the exact same quench implementation and CP-violating term. Computer resources have improved significantly over the last decade and, crucially, we now know that the Higgs mass is 125 GeV. This allows for vastly improved simulations. 

In the following section \ref{sec:model} we recall the model also used in \cite{Tranberg:2006dg}, and briefly present the main observables used to determine the baryon asymmetry. In section \ref{sec:analysis}, we display and comment on our numerical results. We conclude in section \ref{sec:conc}.

\section{The quenched SU(2)-Higgs model with CP-violation}
\label{sec:model}

We model the Higgs-sector of the Standard Model by a Higgs doublet coupled to an SU(2) gauge field, with the classical action
\begin{eqnarray}
S= -\int dt\, d^3x\Bigg[ \frac{1}{2g^2}\textrm{Tr}F^{\mu\nu}F_{\mu\nu} + (D_\mu\phi)^\dagger D_\mu\phi +\mu^2_{\rm eff}(t)\phi^\dagger\phi + \lambda(\phi^\dagger\phi)^2 +\frac{3\delta_{\rm cp}}{16\pi^2 m_W^2}\phi^\dagger\phi \textrm{Tr}F^{\mu\nu}\tilde{F}_{\mu\nu}\Bigg].\nonumber\\
\end{eqnarray}
We have introduced the covariant derivative $D_\mu$ and the field strength $F^{\mu\nu}$ in the usual way, and the last term breaks CP (through breaking P), effectively biasing SU(2) Chern-Simons number,
\begin{eqnarray}
N_{\rm cs}(t)-N_{\rm cs}(0) = \frac{1}{16\pi^2}\int_0^t dt\, d^3x\, \textrm{Tr} F^{\mu\nu}\tilde{F}_{\mu\nu}.
\end{eqnarray}
Chern-Simons number is in turn related through the chiral anomaly to the net baryon number
\begin{eqnarray}
N_B(t)-N_B(0) = 3\left[N_{\rm cs}(t)-N_{\rm cs}(0)\right].
\end{eqnarray}
In our simulations, we will not include fermions (see however \cite{Mou:2015aia}), but infer the baryon asymmetry from the final value of $N_{\rm cs}$. In fact, we will further infer this from the winding number of the Higgs field
\begin{eqnarray}
N_{\rm w}= \frac{1}{24\pi^2}\int d^3x \epsilon_{ijk}\textrm{Tr}[U^\dagger\partial_i UU^\dagger\partial_j UU^\dagger\partial_k U], 
\end{eqnarray}
with 
\begin{eqnarray}
U=\frac{1}{\phi^\dagger\phi} (i\tau_2\phi^*,\phi),
\end{eqnarray}
the normalized matrix representation of the Higgs field.

In contrast to the Chern-Simons number, the Higgs winding number is always integer, and the two coincide at late times, when the gauge and Higgs field configurations are close to pure-gauge. The advantage of using the Higgs winding number is that it settles early in the simulation and is numerically a very "clean" observable, while the Chern-Simons travels around for longer before eventually adjusting to the same final value\footnote{In these simulations, the Chern-Simons number does not suffer from the UV problems inherent to equilibrium computations of the Sphaleron rate \cite{Moore:1998swa,DOnofrio:2014rug}, since equilibrium is not reached and the UV modes never populated. Cooling of the configurations is therefore not required for a reliable calculation.}.

Additional observables that will be of interest include the average Higgs field 
\begin{eqnarray}
\bar{\phi}^2=\frac{1}{V}\int dx^3\phi^\dagger\phi,
\end{eqnarray}
and the total energy. The latter is not conserved as a result of the time-dependent $\mu^2_{\rm eff}(t)$, but decreases in time, more for slower quenches
\begin{eqnarray}
\partial_tE = \frac{d\mu^2_{\rm eff}(t)}{dt}\int d^3x\,\phi^\dagger\phi(x,t), \qquad \Delta E = -\frac{2\mu^2}{\tau_q}\int_0^{\tau_q} dt\, d^3x\,\phi^\dagger\phi(x,t).
\end{eqnarray}
For the slowest quench we consider here, $m_H\tau_q=64$, the total energy is reduced by a factor of about 4, meaning that the final reheat temperature is lower by $4^{1/4}\simeq 1.4$, 32 GeV instead of 45 GeV. However, in a complete model, the quench will be driven by another dynamical degree of freedom (like a scalar field), and the total energy will be conserved. 

We derive the classical equations of motion by variation of the action, and solve them numerically without further approximation. The initial condition is the vacuum state of the Higgs field, when the Higgs potential is simply
\begin{eqnarray}
V_{\rm in}[\phi]=\mu^2\phi^\dagger\phi,
\end{eqnarray}
and where the field and momentum correlators are each given their zero-point fluctuations, sometimes referred to as the "just the half" initialisation \cite{Rajantie:2000nj,GarciaBellido:2002aj,Smit:2002yg}. We initialise only the unstable modes $|k|\leq \mu$. The gauge fields $A_i$ are set to zero initially, with the gauge field momenta $E_{i}$ solved for from Gauss Law in the background of the initial Higgs field. Throughout, we choose the partial gauge fixing $A_0=0$, temporal gauge. We generate an ensemble of random initial configurations and average over the results. The ensemble is engineered to be CP-symmetric, in that for every configuration, we also include the CP-conjugate configuration. Then by construction the asymmetry is exactly zero for $\delta_{\rm cp}=0$.

\section{Results and analysis}
\label{sec:analysis}

\begin{figure}
\centering
    \includegraphics[width = 0.45\textwidth]{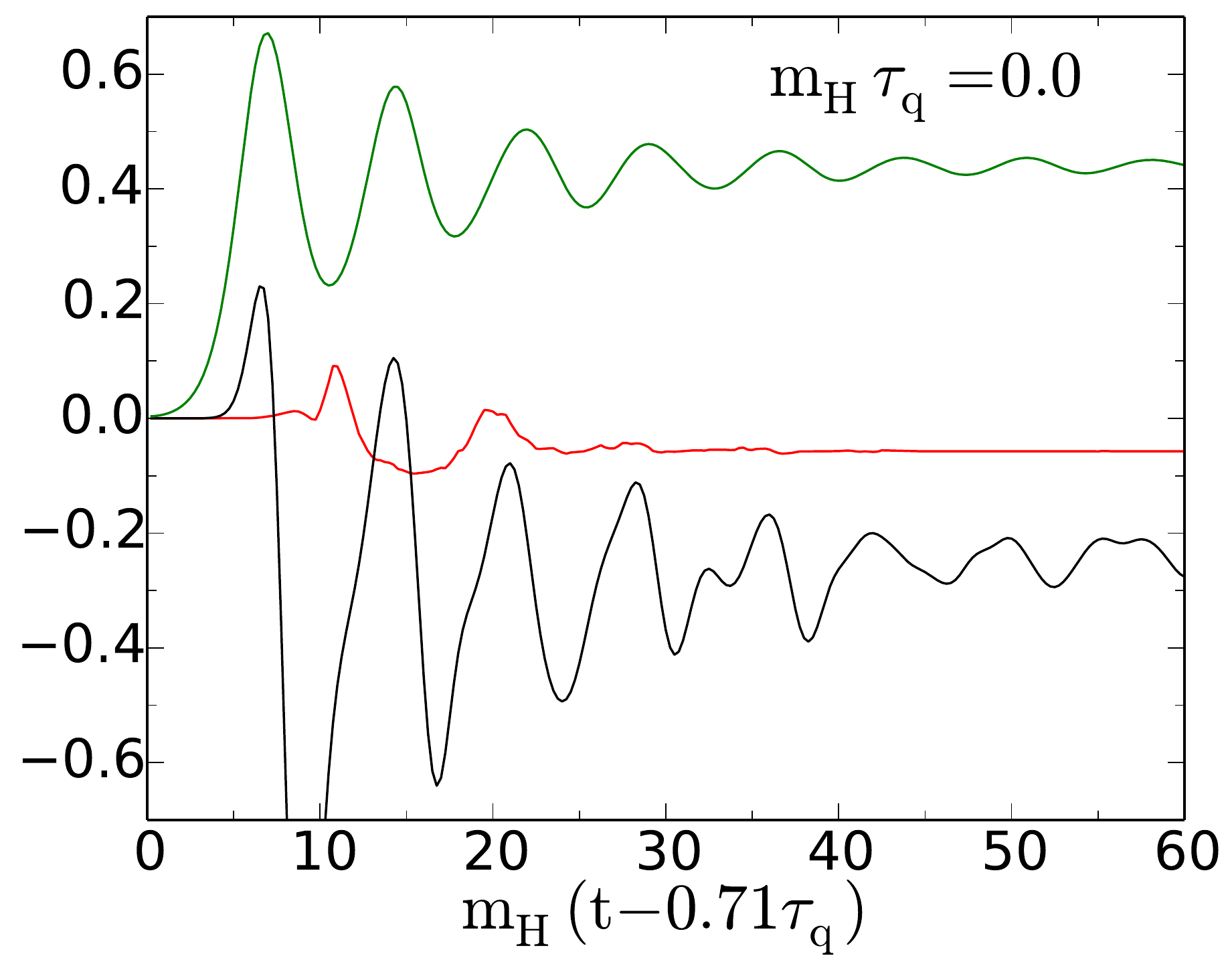}
    \includegraphics[width = 0.45\textwidth]{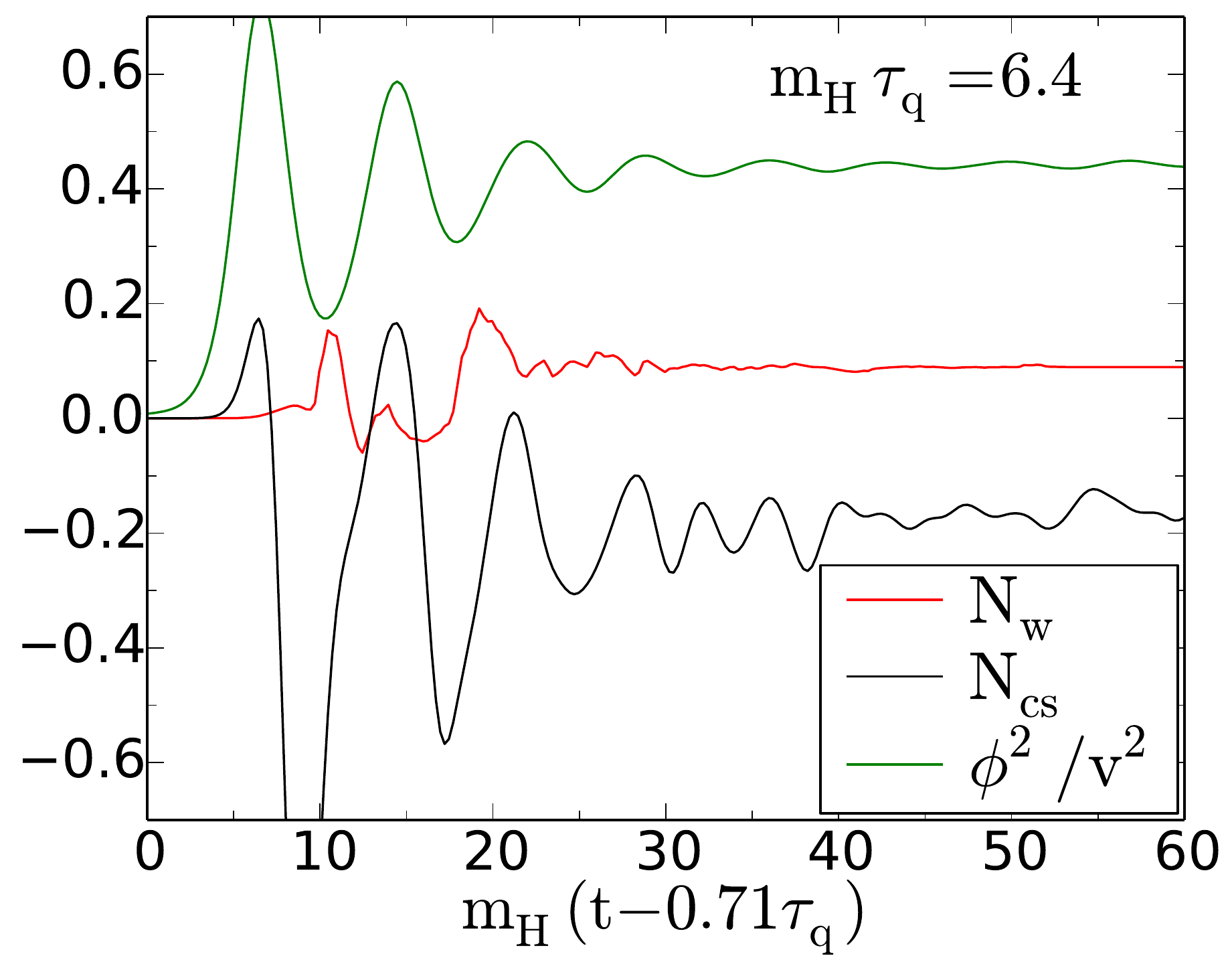}
\caption{The basic observables, ensemble averaged. For quench time $m_H\tau_q=0$ and $6.4$.}
 \label{fig:example1}
\end{figure}

\begin{figure}
\centering
    \includegraphics[width = 0.45\textwidth]{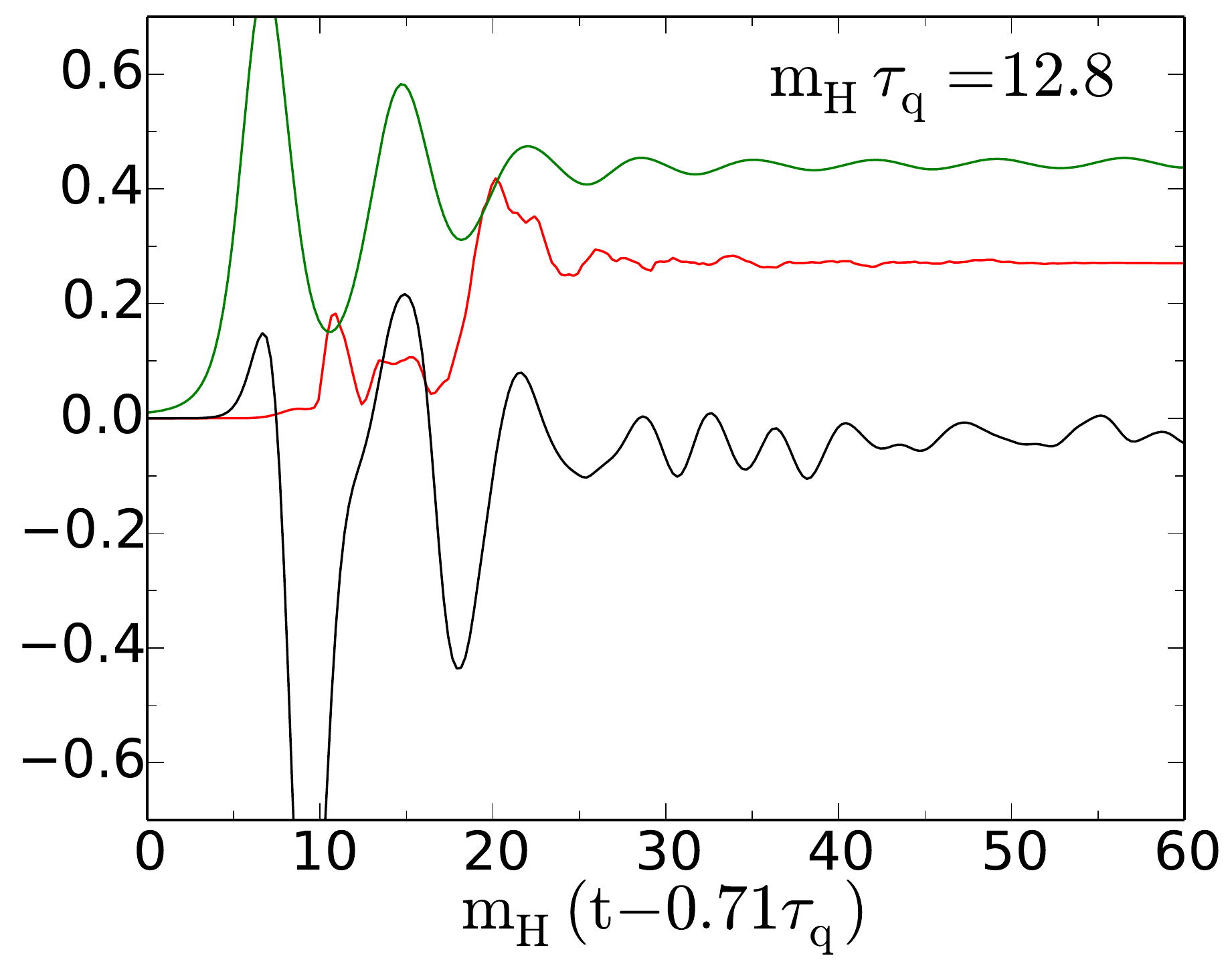}
    \includegraphics[width = 0.45\textwidth]{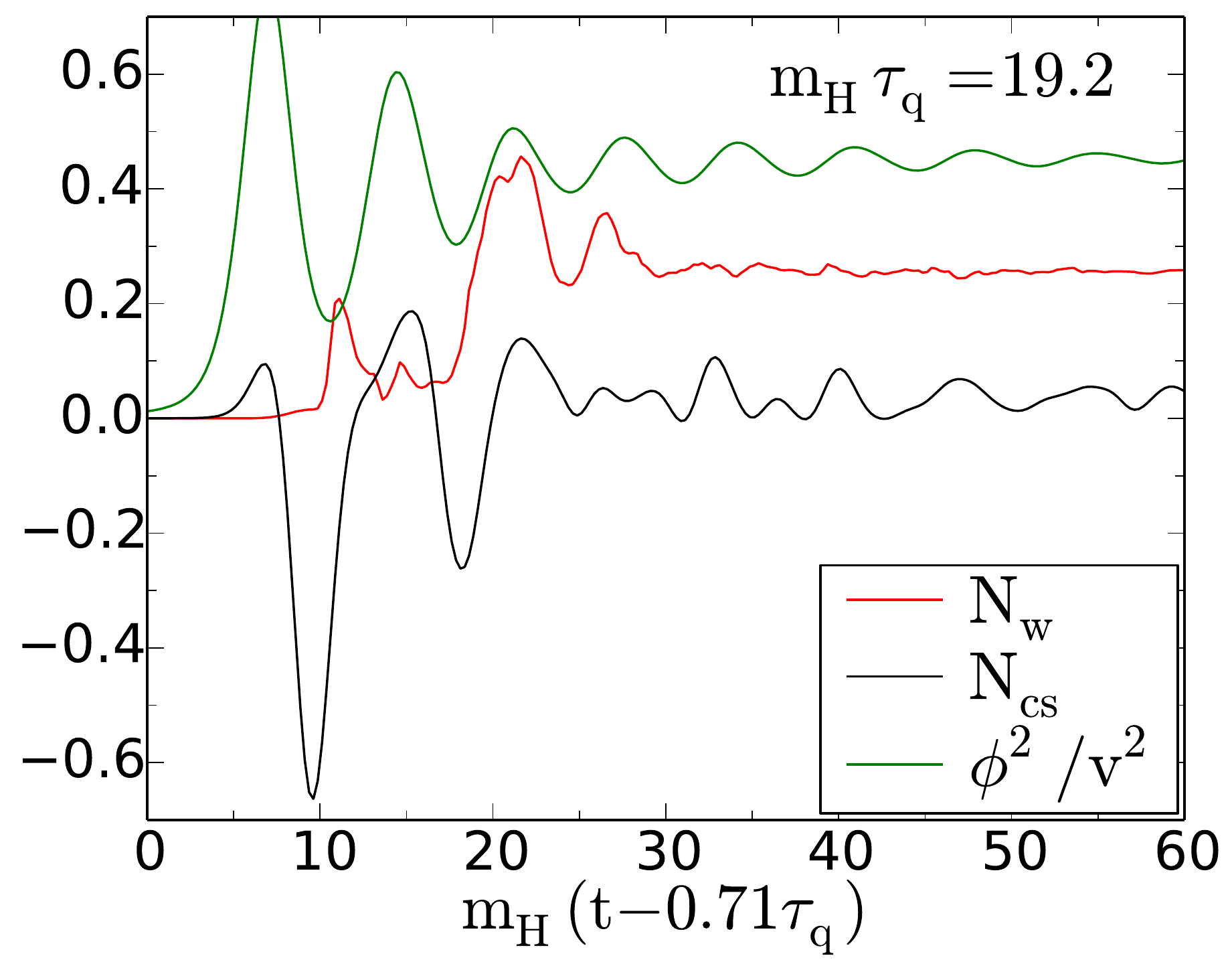}        
\caption{The basic observables, ensemble averaged. For quench time $m_H\tau_q=12.8$ and $19.2$.}
\label{fig:example2}
\end{figure}

\begin{figure}
\centering
    \includegraphics[width = 0.45\textwidth]{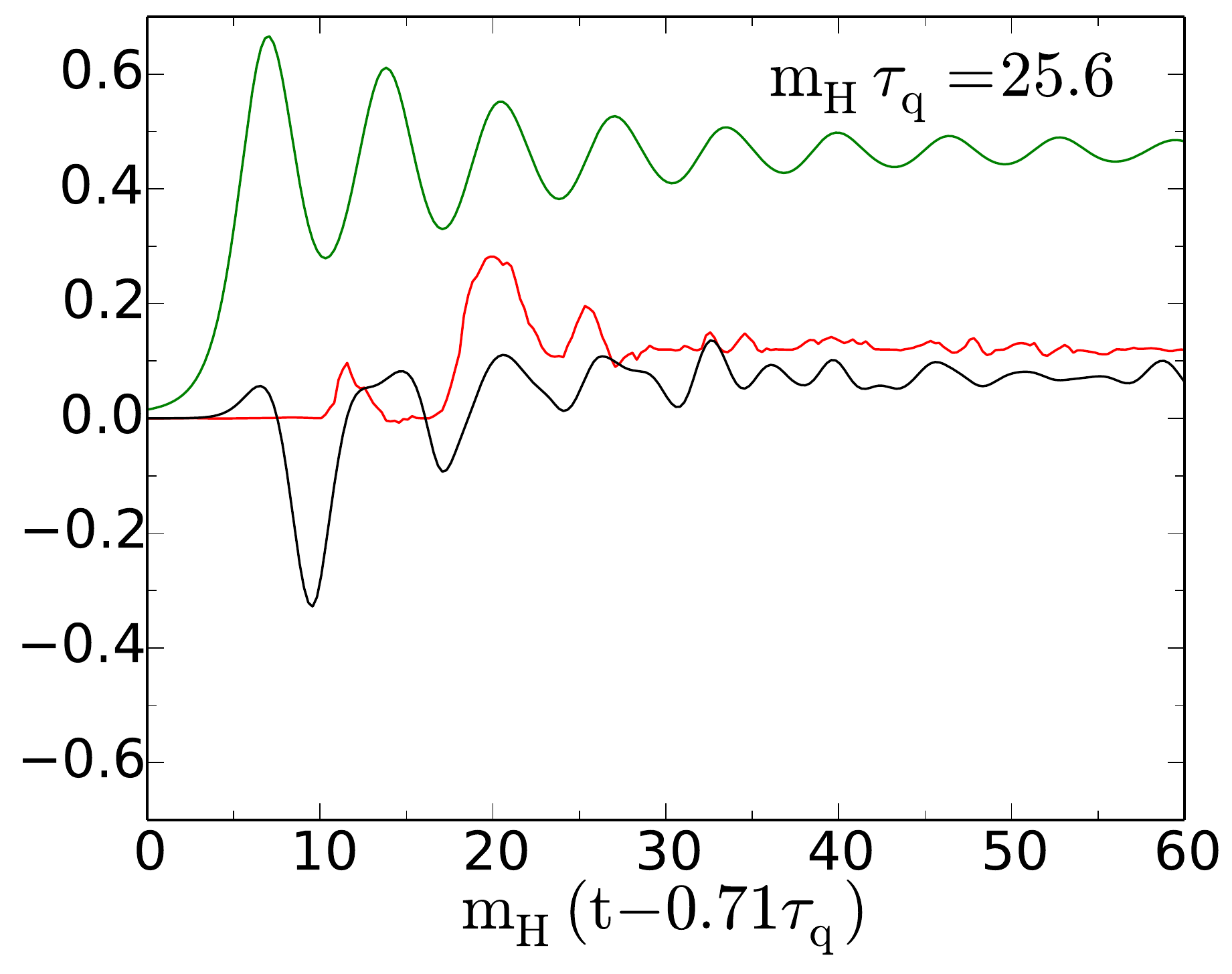}
    \includegraphics[width = 0.45\textwidth]{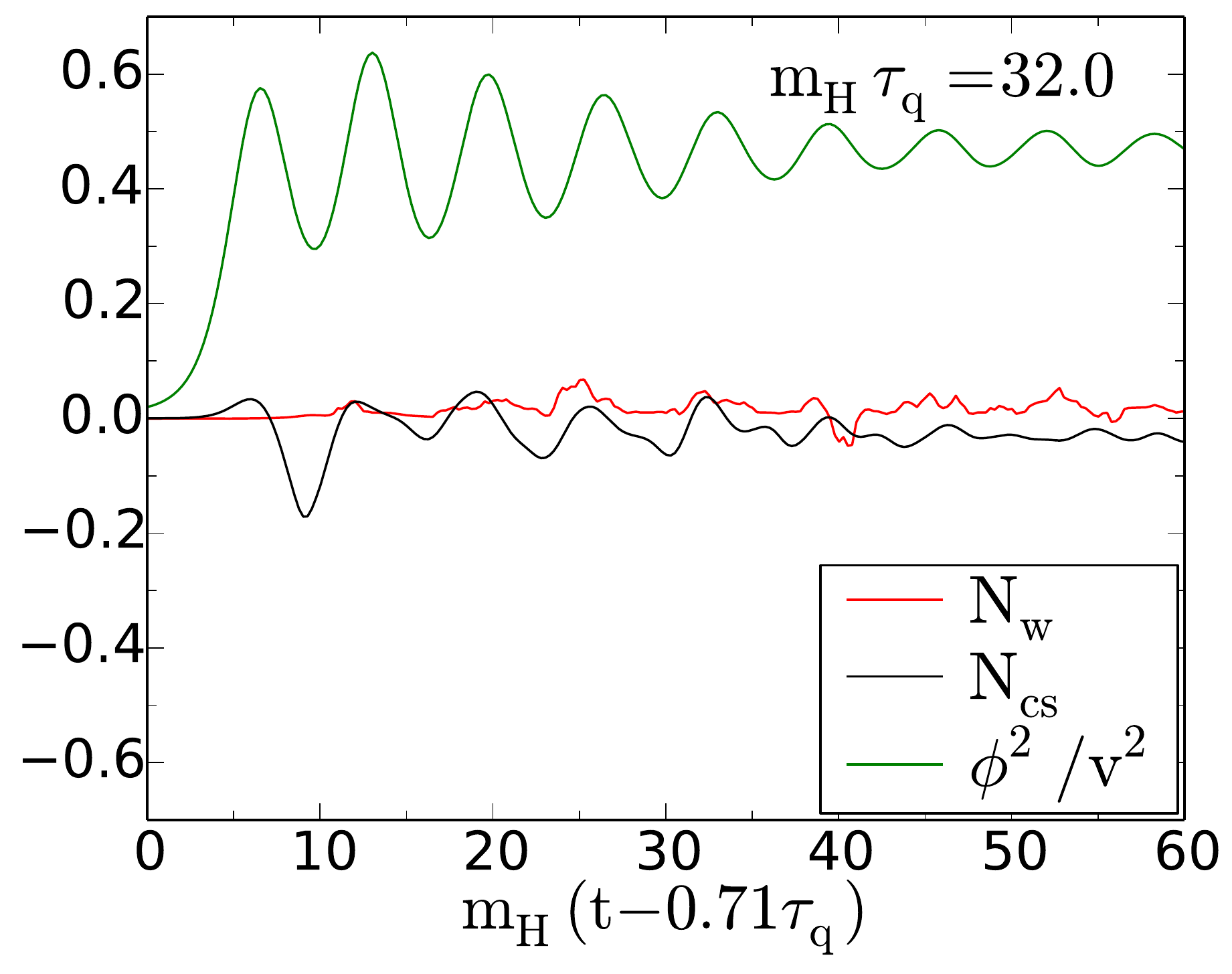}
\caption{The basic observables, ensemble averaged. For quench time $m_H\tau_q=25.6$ and $32$.}
    \label{fig:example3}
\end{figure}

In Figs. \ref{fig:example1}, \ref{fig:example2} and \ref{fig:example3}, we show the behaviour of the main observables $\bar{\phi}^2$, $N_{\rm cs}$ and $N_{\rm w}$ averaged over an ensemble of a few hundred pairs, for a sequence of quench times.  Our lattices have the size $V=(Lm_H)^3=(64\times 0.375)^3$. We see that the Higgs field rolls down the potential and performs a damped oscillation. The Higgs field is not homogeneous, and in fact a large number of zeros of the Higgs field appear at the first few minima of the oscillation \cite{vanderMeulen:2005sp}. These work as nucleation points for potential winding number change. Indeed, we notice that most of the change in $N_{\rm w}$ happens at the Higgs oscillation minima, and that the final asymmetry is generated in the first, second and third Higgs minimum, settling shortly afterwards. Further inspection reveals, that the value of the Higgs field at its first minimum correlates strongly with the final asymmetry. Average Chern-Simons number is violently oscillating, and only much later does it settle to the same value as the winding number (not shown here). 

\begin{figure}
\centering
        \includegraphics[width = 0.55\textwidth]{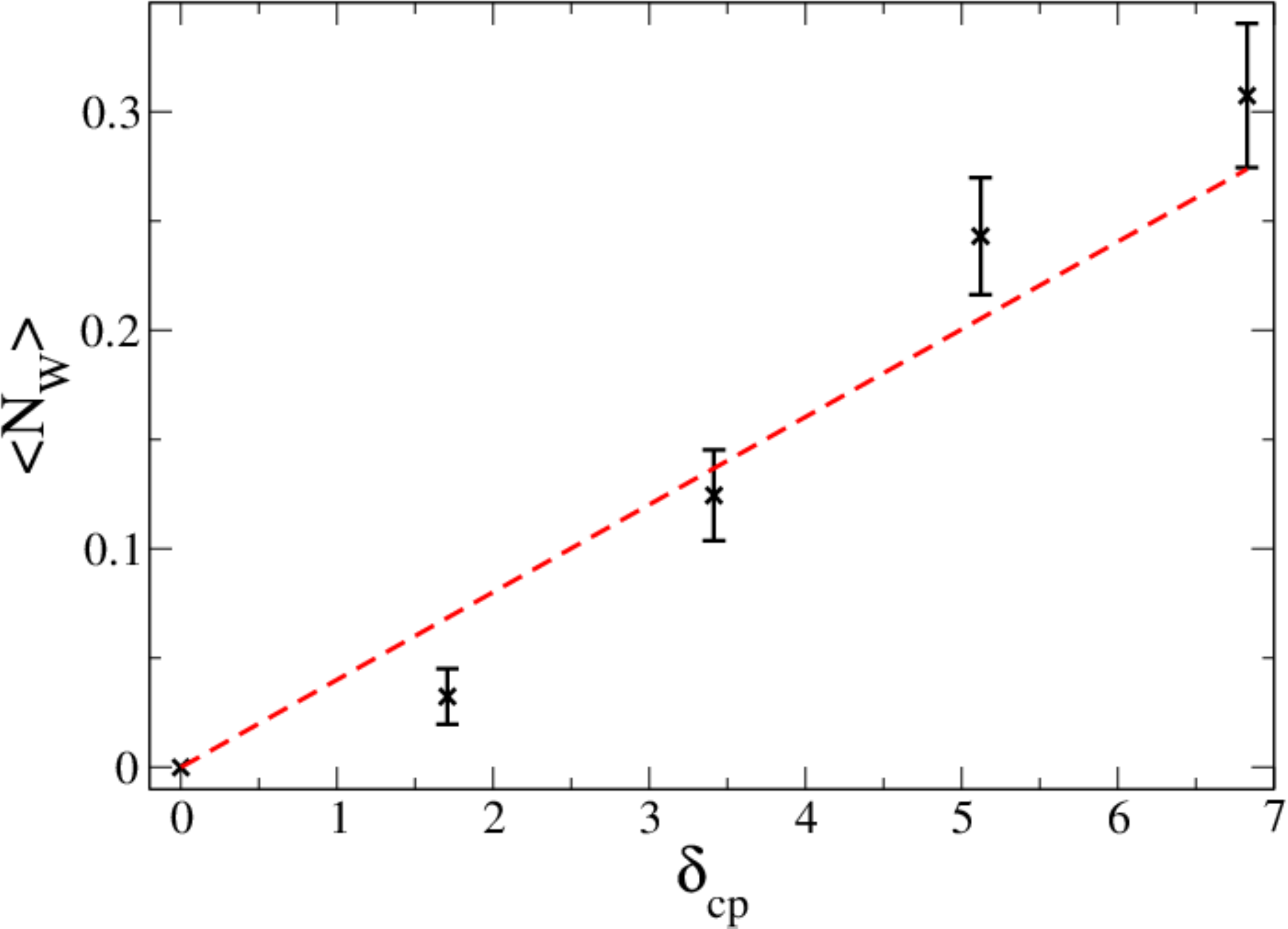}
    \label{fig:kdep}
\caption{The dependence on CP-violation strength for $m_H\tau_q=16$. Overlaid, a linear fit.}
\end{figure}

Given the average winding number, we reconstruct the baryon asymmetry by distributing the total energy in the initial Higgs potential into a thermal final state at temperature $T$, including all the Standard Model relativistic degrees of freedom, $g^*=86.25$. We then find
\begin{eqnarray}
\eta = \frac{n_B}{n_\gamma} = 7.04 \frac{3\langle N_{\rm w}\rangle}{(Lm_H)^3}\frac{45}{2\pi^2 g^*T^3}, \qquad \frac{m_H^4}{16\lambda}= \frac{\pi^2}{30}g^*T^4.
\end{eqnarray}
Given $Lm_H=24$ and a Higgs mass of 125 GeV, we get $\eta = 8.55\times10^{-4} \times\langle N_{\rm w}\rangle$.

We then compute the dependence of the asymmetry on the coefficient $\delta_{\rm cp}$. This was found in \cite{Tranberg:2006ip} to be linear in a range up to at least $\delta_{\rm cp}=1$, for $m_H=2\,m_W$. In Fig. \ref{fig:kdep} we confirm this linear behaviour for quenchtime $m_H\tau_q=16$ up to $\delta_{\rm cp}=7$, now for the physical Higgs mass. Since the asymmetry is odd in $\delta_{\rm cp}$, the next order correction would be $\delta_{\rm cp}^3$, which we found does not improve the fit. Ultimately, in order to match to the observed baryon asymmetry, we will need to interpolate to values very close to zero.  We will employ the linear fit, whereby
\begin{eqnarray}
\eta = 8.55\times 10^{-4}\times(0.040\pm 0.006)\times \delta_{\rm cp} = (3.4\pm 0.5)\times  10^{-5} \delta_{\rm cp}.
\label{eq:kdep_est}
\end{eqnarray}
consistent with \cite{Tranberg:2006ip}.

The Chern-Simons number is biased by the CP-violation term, and the initial rise and subsequent dip (as seen in Figs.~\ref{fig:example1}, \ref{fig:example2}, \ref{fig:example3}) is consistent with a linear response treatment \cite{Rajantie:2000nj,Tranberg:2006dg}. The subsequent violent oscillation are less easy to model. The final asymmetry is an interplay between the dynamical components of the system, many of which a correlated: The availability of winding nucleation points (Higgs field is locally close to zero), energy considerations favouring $N_{\rm cs}\simeq N_W$ and the driving CP-violating force, which may be rewritten by partial integration as 
\begin{eqnarray}
\frac{1}{16\pi^2}\int dt \,d^3x\, \phi^{\dagger}\phi \,\textrm{Tr}F^{\mu\nu}\tilde{F}_{\mu\nu}\propto - \int d^3x \,dt \,\partial_t(\phi^\dagger\phi) n_{\rm cs},
\end{eqnarray}
with $n_{\rm cs}$ the Chern-Simons number density. Hence the driving force is proportional to the speed of the Higgs field, shifted in phase relative to the oscillations producing the Higgs minima. Finally, there is a frequency of the oscillation of the Chern-Simons number itself, related to the boson mass $m_W$. In \cite{Tranberg:2006ip}, it was argued that the non-trivial dependence on Higgs mass could be ascribed to a resonance phenomenon, but here we fix this mass at its physical value.

\begin{figure}
\centering
        \includegraphics[width = 0.65\textwidth]{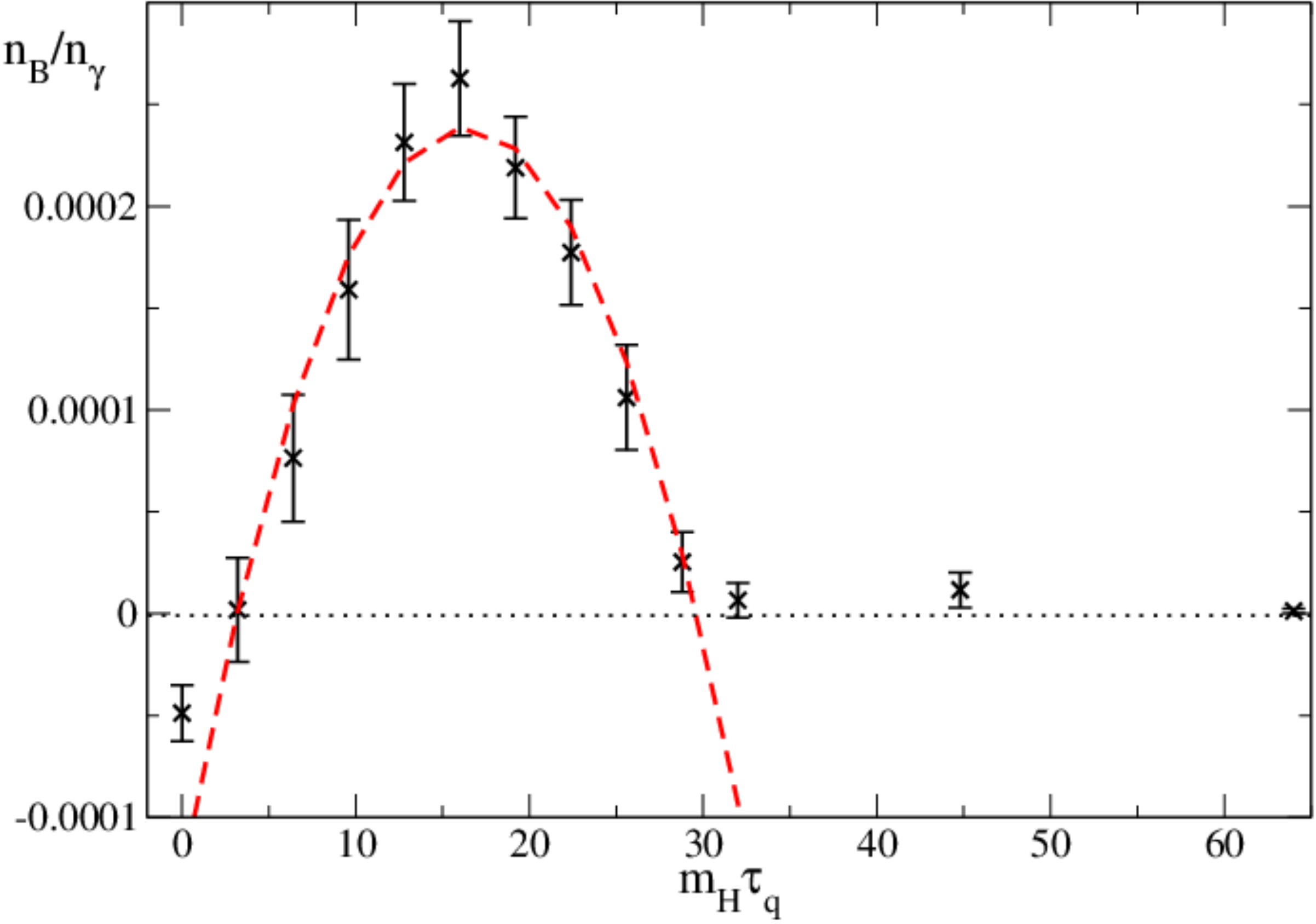}
\caption{The quench time dependence of the baryon asymmetry, for $\delta_{\rm cp}=6.83$.}
 \label{fig:tdep}
\end{figure}

In Fig.~\ref{fig:tdep}, we show our complete results of simulations at different quench times, using $\delta_{\rm cp}=6.83$. We observe that there is a maximum at a finite quench time. Having no theoretical basis for a more specific ansatz, we have fitted the peak with a quadratic form
\begin{eqnarray}
\frac{n_B}{n_\gamma} = A - B(m_H\tau_q-m_H\tau_{\rm max})^2,
\end{eqnarray}
to find using the fitting range $m_H\tau_q\in [6;30]$,
\begin{eqnarray}
m_H\tau_{\rm max}= 16.4\pm 0.2,\qquad A=(3.5\pm 0.1)\times 10^{-5}\delta_{\rm cp},\qquad B=(2.0\pm 0.1)\times 10^{-7}\delta_{\rm cp}.\nonumber\\
\end{eqnarray}
Quench times $m_H\tau_q>30$ produce no asymmetry. Interestingly, the fastest quench\\ $m_H\tau_q=0$ gives a six times smaller asymmetry than the maximum value, with the opposite sign. 

\section{Conclusion}
\label{sec:conc}

Cold Electroweak Baryogenesis may have taken place in the Early Universe, if the Higgs potential was stabillized by interactions with other fields. Then electroweak symmetry breaking could have been delayed until the temperature in the Universe was a few GeV or lower. Alternatively, inflation could have ended at the electroweak scale, with the Universe never reheating above 100 GeV. Many different realisations of such a scenario are possible, involving one or more additional fields. These may or may not be identified as the inflaton, the curvaton, a second Higgs field, a Dark Matter candidate and even composite degrees of freedom.  In order to separate the baryogenesis process from the higher-scale physics of the specific extension of the Standard Model, it is worthwhile computing the generic quench-time dependence of the baryon asymmetry. This allows in the simplest way to match to a specific model.  
In this paper, we pinned down this quench time dependence, showing that there is a preferred value around 
\begin{eqnarray}
m_H\tau_q\simeq 16,\quad (u\simeq 0.09),
\end{eqnarray} 
where the asymmetry is largest and positive. In contrast the fastest quenches produce somewhat smaller asymmetry, potentially of the opposite sign. Quench times longer than twice the optimal value, $m_H\tau_q\geq 30$ ($u<0.05$) give no asymmetry at all. The value of the asymmetry is maximally 
\begin{eqnarray}
\eta =(3.4\pm0.5)\times 10^{-5}\delta_{\rm cp}, 
\end{eqnarray}
so that in this particular implementation of CP-violation, we require $\delta_{\rm cp}\geq 1.8\times 10^{-5}$ to reproduce the observed baryon asymmetry in the Universe of $\eta\simeq 6\times 10^{-10}$.
The creation of the asymmetry is closely associated with the appearance of local zeros of the Higgs field, during its first few oscillations. We have checked that the number of Higgs zeros is not dependent on CP-violation being present, but the CP-bias is most effective at "flipping" the winding and Chern-Simons number near such zeros. It is therefore possible that other sources of CP-violation will exhibit this behaviour, so that our result is more generic than the explicit CP-violating term would suggest. 

It would also be useful to understand the role of the $U(1)$ gauge field of the Standard Model \cite{DiazGil:2007dy,DiazGil:2008tf}. Although it does not in itself enter in the baryon number computation through the anomaly, it may influence the behaviour the system as a whole. Finally, specific implementations of an additional dynamical field should be investigated, in order to establish whether the "by-hand" non-dynamical mass-flip employed here is a good representation of an underlying dynamical system \cite{GarciaBellido:2003wd}.  This work is currently in progress. 

\vspace{0.2cm}

\noindent
{\bf Acknowledgments:}  AT and ZGM are supported by a  UiS-ToppForsk grant from the University of Stavanger. PS acknowledges support by STFC under grant ST/L000393/1. The numerical work was performed on 
on the Abel Cluster, owned by the University of Oslo and the Norwegian metacenter for High Performance Computing (NOTUR), and operated by the Department for Research Computing at USIT, the University of Oslo IT-department.

\end{document}